\pdfoutput=1
\documentclass[11pt,a4paper]{article}
\usepackage{geometry}
\usepackage{graphicx}
\usepackage{color}
\usepackage[colorlinks=true,citecolor=blue,linkcolor=magenta,urlcolor=magenta]{hyperref}
\usepackage[square,numbers,comma,colon,sort&compress]{natbib}
\usepackage{subcaption}
\usepackage{amssymb}
\usepackage{amscd}
\usepackage{amsmath}
\usepackage{tikz}
\usetikzlibrary{patterns,intersections,calc,decorations.markings}

\definecolor{MS}{rgb}{1,0,0}
\newcommand{\braket}[1]{\ensuremath{\langle#1\rangle}}

\begin{document}
\begin{titlepage}

{\hbox to\hsize{\hfill \today }}

\bigskip \vspace{3\baselineskip}

\begin{center}
{\bf \LARGE 
Mechanism for dark matter depopulation }

\bigskip

\bigskip

{\bf Archil Kobakhidze, Michael A.~Schmidt and Matthew Talia \\ }

\smallskip

{ \small \it
ARC Centre of Excellence for Particle Physics at the Terascale, \\
School of Physics, The University of Sydney, NSW 2006, Australia, \\
E-mails: archil.kobakhidze, michael.schmidt, matthew.talia@sydney.edu.au 
\\}

\bigskip
 
\bigskip

\bigskip

{\large \bf Abstract}

\end{center}
\noindent 
Early decoupling of thermally produced dark matter particles due to feeble interactions with the surrounding plasma typically results in their excessive abundance. In this work we propose a simple mechanism for dark matter depopulation. It relies on a specific cosmological evolution under which dark matter particles become temporarily unstable and hence decay away reducing the overall abundance. The instability phase may be followed by an incomplete regeneration phase until the final abundance is established. We explicitly demonstrate this mechanism within a simple toy model of fermionic dark matter and discuss how it can be implemented in theoretically well motivated theories, such as the minimal supersymmetric standard model (MSSM) and for fermionic dark matter in the scotogenic model.

 \end{titlepage}

\section{Introduction}

Stable weakly interacting massive particles (WIMPs) are the most compelling
dark matter (DM) candidates in several extensions of the standard model, notably in
supersymmetric models. Perhaps the most attractive aspect of the WIMP paradigm
is the fact that hypothetical $\sim 100$ GeV stable neutral particles,
interacting with the known standard model particles with the strength of weak
interactions and being once in chemical equilibrium with them, would populate
the universe in abundance comparable to the observed DM abundance.
However, recent null results from direct detection experiments including LUX
\cite{Akerib:2016vxi}, PandaX-II~\cite{Cui:2017nnn} and XENON1T \cite{Aprile:2017iyp} put further constraints on the WIMP interaction cross section. This, along with the absence of any credible evidence for any new physics at the Large Hadron Collider (LHC), has cast doubt on the simplest WIMP hypothesis. Low WIMP cross
section inevitably results in the earlier WIMP decoupling from the primordial
thermal plasma and may lead to an excessive abundance of WIMP dark matter. 

This situation is not atypical. Many well-motivated theoretical models predict over-abundant dark matter. 
For example, bino-like dark matter within the MSSM is over-abundant for most of the parameter area, except the cases when the bino is nearly degenerate in mass with the sleptons.  Another example is fermionic dark matter in the scotogenic model~\cite{Ma:2006km}, where neutrino masses and lepton-flavor-violating processes constrain DM interactions to be small and thus lead to the early decoupling of dark matter in excessive amount.  

The above ramification of the early WIMP decoupling can be altered if the WIMP
temporarily becomes unstable and its abundance is reduced through its
decays.\footnote{Thermal dark matter abundance is also reduced through
co-annihilation and resonant effects \cite{Griest:1990kh}, or through
multi-body scatterings \cite{Dolgov:1980uu, Dey:2016qgf, Choi:2017mkk,
Cline:2017tka, Bell:2017irk}.  These mechanisms typically require a peculiar
compressed spectrum of particle masses.} This could happen if the symmetry
that ensures the stability of the WIMP gets spontaneously broken at some early
cosmological era and is restored at later times.\footnote{A late-time phase transition can also modify the properties of DM, its mass and couplings, during freeze-out.~\cite{Cohen:2008nb}} A similar idea has been
recently put forward in Ref.~\cite{Baker:2016xzo}. The authors of this paper
have not specified the production mechanism for dark matter and presented a
rather contrived model which requires a specific mass arrangement between the
different fields involved. On the contrary, we concentrate on thermal WIMP
models. In addition, the instability phase may be followed
by a regeneration phase, where DM is partially regenerated by the decay of
fields which have been thermally produced during the instability phase.  Our
mechanism of depopulation can be employed in a variety of theoretical frameworks
as is discussed below. Since we are primarily interested in a qualitative
picture in this work, we rely on approximate analytic solutions as opposed to a
comprehensive numerical study of the models.  

The paper is organized as follows. In the next section we discuss the depopulation
mechanism for a generic DM model due to the existence of a DM instability phase in the early universe. In section \ref{sec:regeneration} we discuss the phase of DM regeneration and apply the proposed
mechanism to a simple specific simple model of fermionic DM in section \ref{sec:simplemodel}. In the last section \ref{sec:UVmodels} we briefly discuss how the depopulation mechanism can be implemented within the MSSM and the scotogenic model and draw our conclusions. 

\section{Dark matter depopulation: general consideration}
\label{sec:depop}
Consider generic DM particles with mass $m_{\chi}$ that undergo
scatterings off the standard model particles in the primordial
plasma at some high temperatures. The scattering cross section is assumed
strong enough to keep DM particles in thermal and chemical equilibrium
in the early universe. For the evolution of the cold DM abundance one typically assumes that DM particles become non-relativistic before their decoupling. If DM particles become non-relativistic after the decoupling DM is generically over-abundant. Since our scenario involves a phase where the DM abundance gets reduced the assumption of non-relativistic DM decoupling is not necessary. However, we must ensure that DM becomes non-relativistic during the depopulation phase to avoid DM repopulation due to inverse decays. 

The DM abundance yield $Y_{\chi}(x)=n_{\chi}/s$ ($n_{\chi}$ and $s$ are
particle number and entropy densities, respectively), evolves according to
the following Boltzmann equation:
\begin{equation}
	\frac{dY_{\chi}}{dx}=-\frac{\lambda}{x^{n+2}}\left(Y_{\chi}^2-Y_{\chi}^{(eq)2}\right)~
\label{1a}
\end{equation} 
where $x\equiv m_{\chi}/T$, $m_{\chi}$ being the DM particle mass, and
\begin{equation}
	\lambda= \left[\frac{x s_{\chi}\braket{\sigma v}}{H_{\chi}}\right]_{x=1} \times \begin{cases}
		1 & \text{$\chi$ is its own antiparticle}\\
		\frac12 & \text{otherwise}\\
	\end{cases}\;.
\label{2a}
\end{equation}
In the above equation $s_{\chi}=(2\pi^2/45)g_{*}m_{\chi}^3$ and
$H_{\chi}=(\pi^2g_{*}/90)^{1/2}\frac{m_{\chi}^2}{M_P}$ are, respectively, the
entropy density and the Hubble expansion rate both evaluated at $T=m_{\chi}$.
$M_P\approx 2.44\times 10^{18}$ GeV is the reduced Planck mass and $g_{*}$
denotes the number of relativistic degrees of freedom in thermal equilibrium
at a given temperature. 
For our purpose it is safe to ignore changes in $g_{*}$ with temperature. We
also approximate the thermally-averaged cross section $\braket{\sigma v}$ by its leading partial wave and thus regard $\lambda$ as an $x$-independent constant. The power of $x$ is given by $n=0$ for $s$-wave scattering and $n=1$ for $p$-wave scattering. 

Initially, $x \lesssim {\cal O}(1)$, the number density is approximately equal to
the equilibrium number density $Y_{\chi}(x)=Y^{(eq)}_{\chi}(x)+\delta \approx
Y^{(eq)}_{\chi}(x)$, where $\delta$ measures small departure from the equilibrium
abundance yield. The equilibrium yield  $Y^{(eq)}_{\chi}(x)$ is: 
\begin{eqnarray}
Y_{\chi}^{(eq)}=\left\lbrace
\begin{array}{ll}
	\frac{45\zeta(3)}{2\pi^4}\frac{g_{\chi}}{g_{*}} \simeq 0.278\frac{g_{\chi}}{g_{*}} & \mathrm{relativistic (bosons)} \\ 
	\\
	\frac34\frac{45\zeta(3)}{2\pi^4}\frac{g_{\chi}}{g_{*}} \simeq 0.208\frac{g_{\chi}}{g_{*}} & \mathrm{relativistic (fermions)} \\ 
\\
\frac{45}{2\pi^4}\left(\frac{\pi}{8}\right)^{1/2}\frac{g_{\chi}}{g_{*}}x^{3/2}{\rm e}^{-x} & \mathrm{non-relativistic}
\end{array} 
\right.
\label{3a}
\end{eqnarray}
where $g_{\chi}$ measures the degrees of freedom per DM particle, e.g.~$g_\chi=\tfrac74$ for a Majorana fermion.  

Note, for relativistic particles $Y_{\chi}^{(eq)}$ stays constant and so is 
$Y_{\chi}(x)$ ($\delta=0$) up until the DM particles become non-relativistic, $x>1$.  The evolution of non-relativistic DM density is more complicated. Namely, for
non-relativistic DM particles $Y_{\chi}(x)$ decreases exponentially together with the equilibrium yield as the universe
cools down. At sufficiently low temperature
 $x_f$ ($x_f\approx 20$ for non-relativistic DM), the DM number density drops to the point when
 scatterings of DM particles become rare and DM is no longer
 able to sustain itself in the thermal bath. Close to this freeze-out
 temperature departure from the equilibrium yield becomes significant and
 $Y_{\chi}\approx \delta \gg Y^{(eq)}_{\chi}(x)$ with $x\sim x_f$ and:
\begin{equation}
	\delta(x)\approx \frac{x^{n+2}}{2\lambda}~,
\label{4a}
\end{equation}
while for relativistic DM still $\delta=0$.  
Subsequently at $x \gg x_f$ and $x \gg 1$,  the non-relativistic equilibrium yield becomes negligibly small and can
be dropped from eq. (\ref{1a}). The analytic solution for the abundance
yield at present time $Y_{\chi,0}\approx Y_{\chi}(x\to \infty)$ then reads as:
\begin{equation}
	Y_{\chi,0}\approx \frac{(n+1) Y_{\chi}(x_f)x_f^{n+1}}{Y_{\chi}(x_f)\lambda+(n+1)x_f^{n+1}}
\label{5a}
\end{equation} 
where $Y_{\chi}(x_f)= \delta(x_f)$ for non-relativistic DM ($x_f>1$) or $Y_{\chi}(x_f)= Y_{\chi}^{(eq)}\simeq 0.208 (0.278) \frac{g_{\chi}}{g_{*}}$ for relativistic fermionic (bosonic) DM are defined by matching the solutions in two different regimes. 
Using the expression in eq. (\ref{4a}) for $\delta (x_f)$ and $x_f\gg 1$ for non-relativistic, eq. (\ref{5a}) we obtain an approximate solution for the DM yield and thus find 
\begin{equation}
Y_{\chi,0}\approx 
\left\lbrace
\begin{array}{ll}
	0.278\frac{g_{\chi}}{g_{*}} & \mathrm{relativistic (bosons)} \\ 
\\
	0.208\frac{g_{\chi}}{g_{*}} & \mathrm{relativistic (fermions)} \\ 
\\
\frac{n+1}{\lambda} x_f^{n+1} & \mathrm{non-relativistic}
\end{array} 
\right.
\label{6a}
\end{equation}
Since typically $\lambda \gg 1$, the DM abundance of relativistically decoupled DM is typically several orders of magnitude higher than that of non-relativistically decoupled DM.  

Now let us assume that there is a cosmological phase transition at $x_a$ to a phase where DM becomes unstable due to the spontaneous breaking of the symmetry which stabilizes DM, followed by
the restoration of this symmetry at $x_b$. In this instability phase $x\in
[x_a,x_b]$ the DM particle is allowed to decay. As we have mentioned previously DM must be non-relativistic during at least part of this instability phase to ensure that inverse decays are suppressed and thus $x_b>1$. The Boltzmann equation
then gets modified as:
\begin{equation}
	\frac{dY_{\chi}}{dx}=-\frac{\braket{\Gamma_{\chi}}x}{H_{\chi}}\left(Y_{\chi}-Y_{\chi}^{(eq)}\right)-\frac{\lambda}{x^{n+2}}\left(Y_{\chi}^2-Y_{\chi}^{(eq)2}\right) 
	- Y_\chi \frac{d\ln S}{dx}
	\;,
\label{7a}
\end{equation}    
where $n=0$ for $s$-wave annihilation, $n=1$ for $p$-wave
annihilation. Here $\braket{\Gamma_{\chi}}=\Gamma_\chi K_1(x)/K_2(x)$ is the thermally-averaged DM particle decay width with the zero temperature decay width $\Gamma_\chi$ and the modified Bessel functions $K_n(x)$.  For non-relativistic DM it is approximately given by the zero temperature decay width, since $K_1(x)/K_2(x)=1-3/(2x)+\mathcal{O}(x^{-2})$ for $x\gg 1$, while it is strongly suppressed when DM is relativistic, $K_1(x)/K_2(x)=x/2+\mathcal{O}(x^2)$ for $x\ll1$. As we are mostly interested in the region of parameter space when $\chi$ is (close-to) non-relativistic and to allow for an approximate analytic solution, we approximate it by the zero temperature decay width. If DM is relativistic during the initial stages of the instability phase, an approximate solution is obtained by neglecting DM decay before it becomes non-relativistic.

The last term describes the change in entropy $S$ during the thermal evolution.
The main contributions to entropy production are particle decoupling, but also a first order phase transition may generate additional entropy. 
If the DM yield is close to its equilibrium value, the change in entropy can be parameterized by an entropy dilution factor $S(x)/S(x_0)$ which relates the DM yields as $Y_{\chi}(x) = Y_\chi(x_0) S(x_0)/S(x)$. The same conclusion holds if the DM distribution is much larger than its equilibrium distribution, i.e.~$Y_\chi \gg Y_\chi^{(eq)}$, and annihilations can be neglected compared to decays. In the most general case there is no analytic solution. In the following we assume
that the instability phase occurs well above $T=1$ GeV and
thus we can neglect entropy production from particle
decoupling, i.e.~the number of relativistic degrees of freedom $g_*$ is approximately constant, and we approximate entropy production during a
first order phase transition by an entropy dilution factor. 
Thus in the following discussion we neglect the last term in eq.~\eqref{7a}, but remind the reader that for a first order phase transition entropy dilution factors have to be included in the final solution.

The effect of DM decays on its final abundance critically depends on when the
instability phase took place. If it happened well before the standard
freeze-out, i.e., $x_f \gg x_b$, the subsequent scatterings can repopulate dark
matter particles leading essentially to the standard abundance (\ref{6a}).
However, a dramatic change in the final DM abundance is expected, when
the instability phase follows the freeze-out (out of equilibrium decays), i.e.
$x_f \ll x_a$,  or DM particles freeze-out during ($x_a<x_f<x_b$)  or
just after the instability phase ($x_f\sim x_a$). We separately consider the cases whether the equlibrium abundance of $\chi$ can be neglected or not.

\subsection{Negligible equilibrium abundance}
If the standard freeze-out precedes the instability phase, $x_f\ll x_a$, and the decay rate is sufficiently small such that the equilibrium yield is always negligible, then one can
ignore the equilibrium yield in eq. (\ref{7a}),
which then becomes the Bernoulli
equation and can be solved analytically (in terms of the incomplete $\Gamma$ function):
\begin{align}
	Y_\chi(x_b) &= \frac{ Y_{\chi }(x_a)    e^{-\frac{\Gamma _{\chi }}{2 H_{\chi }} x_b^2}}{ e^{-\frac{\Gamma_{\chi}}{2 H_\chi}x_a^2}  + \lambda\, Y_\chi(x_a)\,\frac{1}{2} \left(\frac{\Gamma _{\chi }}{2 H_{\chi }}\right)^{\frac{n+1}{2}}
	\left[
		\Gamma \left(\frac{-1-n}{2},\frac{ \Gamma _{\chi }}{2 H_{\chi }} x_a^2\right)
 - \Gamma \left(\frac{-1-n}{2},\frac{ \Gamma _{\chi }}{2 H_{\chi }} x_b^2\right)
\right]}\\
&\approx
\frac{  Y_{\chi }(x_a)    e^{-\frac{\Gamma _{\chi }}{2 H_{\chi }} (x_b^2-x_a^2)}}{ 1 + \lambda\,Y_\chi(x_a) \,\frac{  H_\chi}{\Gamma_\chi} 
	\left[
		\frac{1}{x_a^{3+n}}
		-\frac{e^{-\frac{\Gamma_\chi}{2 H_\chi} (x_b^2-x_a^2)}}{x_b^{3+n}}
\right]} \label{8a}
   \end{align}
   We have used an asymptotic expansion of the incomplete $\Gamma$-function in $x$
   to make the approximation in the last line. This approximation is accurate for all
practical purposes. The 2-by-2 scattering cross section enters the solution (\ref{8a})
explicitly only through the matching condition, i.e. $\lambda\, Y_\chi(x_a) \approx (n+1)x_f^{n+1}$, see eq.~\eqref{6a}. We observe an exponential
suppression factor which has clear intuitive explanation: the reduction of DM number density is essentially defined through the decay rate of DM particles per the universe expansion rate and the duration of the instability
phase, $\Delta t \propto (x_b^2-x_a^2)$. The present abundance yield can be
obtained from eq. (\ref{5a}) by the substitution: $Y_{\chi}(x_f)\to
Y_{\chi}(x_b)$ in the absence of any regeneration phase as discussed in the next section. 

Consider now the case where the standard freeze-out follows the instability
phase, i.e. $x_f \sim x_b$. In this case the solution to eq. (\ref{7a}) is
given by $Y_{\chi}(x)=Y_{\chi}^{(eq)}+\delta_d$, where
 \begin{equation}
\delta_d (x)\approx -\frac{{\rm d}Y^{(eq)}_{\chi}}{{\rm d}x}\left[\frac{\Gamma_{\chi}}{H_{\chi}}x+\frac{2\lambda}{x^{n+2}}Y_{\chi}^{(eq)}\right]^{-1}\;.
\label{9a}
\end{equation} 
This must be compared with the standard solution (\ref{4a}). Unless the decay rate $\Gamma_{\chi}$ is extremely small, the first term in the denominator dominates and $Y_{\chi}^{(eq)}(x_f) \gg \delta_d(x_f)$. Hence, unlike scatterings, decays keep the DM abundance exponentially close to its equilibrium value even at the freeze-out temperature. The final yield, therefore, is obtained from eq. (\ref{5a}) by the substitution $Y_{\chi}(x_f)\to Y^{(eq)}_{\chi}(x_f)$ and may be substantially smaller than the standard abundance in eq.~(\ref{6a}).         

Furthermore, consider the case where freeze-out happens during the instability phase, i.e. $x_a < x_f <x_b$. The solution for $x_a<x_f$ is given by eq. (\ref{9a}), while for $x_f<x_b$ the solution is given by (\ref{8a}). We combine these solutions by imposing the matching condition: $Y_{\chi}(x_f)=Y_{\chi}^{(eq)}(x_f)+\delta_d(x_f)$. The present day abundance then is again obtained from eq. (\ref{5a}) via the substitution: $Y_{\chi}(x_f)\to Y_{\chi}(x_b)$. 

\subsection{Sizable equilibrium abundance}
\label{subsec:sizeq}
If inverse decays are fast enough to keep DM in equilibrium, we obtain an approximate analytic solution for the present day abundance from eq. \eqref{5a} by replacing $x_f$ by $x_b$ and $Y_\chi(x_f)$ by the equilibrium abundance $Y_\chi^{(eq)}(x_b)$  
\begin{equation}
	Y_{\chi,0} \approx \frac{(n+1) Y_{\chi}^{(eq)}(x_b) x_b^{n+1}}{Y_\chi^{(eq)}(x_b) \lambda + (n+1) x_b^{n+1}}
\end{equation}
Finally, the intermediate regime where the inverse decays are relevant, but not fast enough to keep DM in equilibrium, requires a numerical solution.


\section{Dark matter regeneration: general consideration}
\label{sec:regeneration}

During the instability phase, $x\in [x_a,x_b]$, the scalar
$S$ which spontaneously breaks the symmetry has a
vanishingly small thermal mass close to the critical
temperature. Similarly there may be Goldstone bosons or
massive gauge bosons from the breaking of additional
continuous symmetries. All these light degrees of freedom
may be produced during the instability phase via decays of
the DM candidate and scatterings with SM particles. The mass of these particles increases during the instability phase with decreasing temperature, because the VEV of the scalar $S$ increases, $\braket{S}^2 \propto 1-x_a^2/x^2$, and thus depending on the second phase transition the light degrees of freedom may stay relativistic during the instability phase or become non-relativistic, if their thermal mass exceeds the temperature before the second phase transition.

Ultimately after the symmetry is restored at $x_b$ all light
degrees of freedom of $S$ are heavy and follow a Maxwell-Boltzmann distribution. Relativistic degrees of freedom of $S$ which become non-relativistic at $x_b$ are far from their equilibrium distribution. We first focus on them and discuss degrees of freedom which become non-relativistic during the instability phase below. We assume that all relativistic degrees of freedom of $S$ are in kinetic and chemical equilibrium following a Bose-Einstein distribution during the instability phase (it is
sufficient if they are at the end).
We assume that the scalars are still in kinetic equilibrium with the SM thermal bath after the phase transition and thus have the same temperature. This is satisfied if the relaxation rate $\Gamma_{relax}\simeq \Gamma_{coll}/N_{coll}$ (expressed in terms of the collision rate $\Gamma_{coll}$ and the number of collisions $N_{coll}$) exceeds the Hubble rate 
$H \lesssim \Gamma_{relax} \simeq \frac{T}{3m_S} \Gamma_{coll}$ which we assume in the following.\footnote{This condition is generally fulfilled, if the scalar interacts via electroweak interactions: The collision rate is $\Gamma_{coll}\simeq G_F^2 T^5$ for a (non-relativistic) particle $S$ scattering with a relativistic particle in the SM thermal bath, and thus kinetic equilibrium is achieved for $T^4\gtrsim G_F^{-2} m_S/M_P \simeq (40\,\mathrm{MeV})^4 (m_S/\mathrm{TeV})$.}
Their number density
does not change during the symmetry-restoring phase
transition, hence we have:
\begin{equation}
	\frac{\zeta(3)}{\pi^2} T_b^3  
	=\left(\frac{m_S T_b}{2\pi}\right)^{3/2} e^{(\mu-m_S)/T_b}  
\qquad	\Rightarrow \qquad
	\mu  \approx
	m_S\;.
\end{equation}
As $\mu\sim m_S\gg T_b$ the scalar $S$ is initially over-abundant, with $Y_S/Y_S^{(eq)}\sim \exp({m_S/T_b})$. 

Particles which became non-relativistic during the instability phase follow a Maxwell-Boltzmann distribution both before and after the second phase transition. As their mass may change during the phase transition, but their number density remains the same during the phase transition, we find 
\begin{equation}
\left(\frac{m_{S}^\prime T_b}{2\pi}\right)^{3/2} e^{-m_{S}^\prime/T_b}  
	=\left(\frac{m_S T_b}{2\pi}\right)^{3/2} e^{(\mu-m_S)/T_b}  
\qquad	\Rightarrow \qquad
	\mu  \approx
m_S-m_{S}^\prime+\frac32 T_b\ln\frac{m_{S}^\prime}{m_S}
\end{equation}
where $m_S^\prime$ denotes the mass of the scalar at the end of the instability phase and $m_S$ the zero temperature mass. Thus the scalar may also be over-abundant, $Y_S/Y_S^{(eq)}\sim \left(\tfrac{m_S^\prime}{m_S}\right)^{3/2} \exp({\tfrac{m_S-m_S^\prime}{T_b}})$, directly after the phase transition, although its abundance is closer to the equilibrium abundance compared to above.

As the scalars are heavier than the DM mass and non-relativistic, they quickly annihilate to SM particles and decay to the DM candidate and a SM particle. In analogy to DM annihilation these processes are described by\footnote{Note that DM annihilations are neglected, because we assume that DM freeze-out occurs before the regeneration phase, i.e.~$x_f<x_b$.}
\begin{align}
	\frac{dY_{S}}{dx} &=-\frac{\Gamma_{S}x}{H_{\chi}}\left(Y_{S}-\frac{Y_\chi}{Y_{\chi}^{(eq)}}Y_{S}^{(eq)}\right)-\frac{\lambda_S}{x^{m+2}}\left(Y_{S}^2-Y_{S}^{(eq)2}\right) \label{eq:S}\\
		\frac{dY_{\chi}}{dx} &=\frac{\Gamma_{S}x}{H_{\chi}}\left(Y_{S}-\frac{Y_\chi}{Y_{\chi}^{(eq)}}Y_{S}^{(eq)}\right)\;,\label{eq:DMprod}
\end{align}
where $\Gamma_S$ is the zero temperature scalar decay width which is a valid approximation in this case for the thermally averaged decay width, since $S$ is non-relativistic. The annihilation cross section of $S$ into SM particles is parameterized by
\begin{equation}
	\lambda_S\equiv \left[ \frac{x s_\chi \braket{v\sigma(SS\to f \bar f) }}{H_\chi}\right]_{x=1} \times\begin{cases} 1 & \text{$S$ is its own antiparticle}\\
		\frac12 & \text{otherwise}
	\end{cases} \;.
\end{equation}
Note that the DM may in general not be in chemical equilibrium, i.e. $Y_\chi\neq Y_\chi^{(eq)}$. 
We again approximate the thermally-averaged cross-section by its leading
partial wave. The power of $x$ is given by $m=0$ for $s$-wave scattering
and $m=1$ for $p$-wave scattering. 

There are several relevant timescales: (i) freeze-out of $S$, $H(x_{f,S}) =
\left.\braket{v\sigma(SS\to f\bar f)} n_S\right|_{x_{f,S}}$, (ii) chemical
	equilibrium of $S$, $Y_S(x_c)=Y_S^{(eq)}(x_c)$ and (iii) the time when
	inverse decays become important, $Y_S(x_i) Y_\chi^{(eq)}(x_i) =
	Y_\chi(x_i) Y_S^{(eq)}(x_i)$. The phenomenology is different depending
	on the relative ordering of the timescales.  As $Y_S(x_b)\gg
	Y_S^{(eq)}(x_b)$ and $Y_S$ is a decreasing function of $x$ for $Y_S(x)>Y_S^{(eq)}(x)$, we generally find $x_i\leq x_c$. 
	
\begin{figure}[t]\centering
	\includegraphics[width=0.6\linewidth]{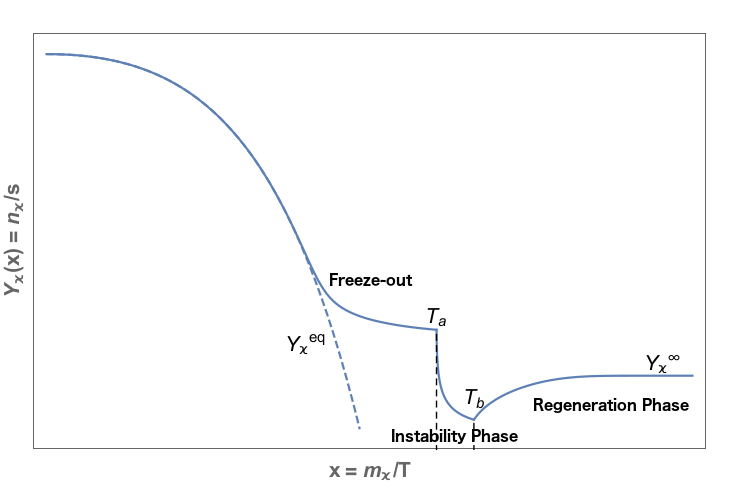}
	\caption{A schematic illustration of the DM abundance where the standard freeze-out is followed by an instability phase where DM decays ($T_a>T>T_b$) and regeneration of DM ($T<T_b$) until the final abundance $Y_{\chi}^{\infty}$ is established. }
	\label{fig1}
\end{figure}

	As inverse decays become important very quickly in the simple example model and thus a numerical solution is required, we do not give an explicit analytic solution, but only state that similar approximations as in the previous section may be used to obtain analytic solutions for some regions of parameter space.
The DM depopulation scenario is schematically summarized on Fig.~\ref{fig1}. This obviously only shows one scenario. DM may also freeze-out during the instability phase or while still being relativistic. 


\section{Simple fermionic model}\label{sec:simplemodel}
We illustrate the mechanism described in the previous section within a simple toy model which contains an additional electroweak scalar doublet $\eta$ and a Majorana fermion $N$ apart from the SM particles. Both $\eta$ and $N$ are odd under an imposed $Z_2$ symmetry ($Z_2:$ $\eta\to -\eta$ and $N\to -N$), while SM particles are $Z_2$-even. We largely follow the notation in Ref.~\cite{Ginzburg:2010wa} and refer the reader to this reference for more details. This model has direct applications to the scotogenic model~\cite{Ma:2006km} with fermionic DM and bino-like DM in the MSSM. 

The most general Lagrangian involving extra fields is given by
\begin{align}
	\mathcal{L}= i N^\dagger \bar\sigma^\mu\partial_\mu N  + (D_\mu \eta)^\dagger D^\mu\eta - V(H,\eta)  - \left(\frac12 M_N N N + Y_{\alpha} N L_\alpha \cdot \eta  +\mathrm{h.c.} \right)
\end{align}
where we chose the Majorana mass term of the SM singlets to be diagonal with 
\begin{align}
	V(H,\eta)  =  &- \frac12 \left[ \mu_H^2 H^\dagger H +\mu_\eta^2 \eta^\dagger\eta \right] +\frac12 \left[ \lambda_1 (H^\dagger H)^2  + \lambda_2 (\eta^\dagger\eta)^2 \right] 
	\\\nonumber &
	+  \lambda_3\, (H^\dagger H) (\eta^\dagger \eta) + \lambda_4\, (H^\dagger \eta) (\eta^\dagger H) +\frac{\lambda_5}{2}\left[\left(H^\dagger \eta\right)^2+ \left(\eta^\dagger H\right)^2\right]
\end{align}
with real couplings $\lambda_i$ and $\mu_i^2$ and use the freedom of phase redefinitions of $H$ and $\eta$ to choose $\lambda_5<0$. We focus on the case with $\mu_\eta^2>0$\footnote{The alternative scenario with $\mu_\eta^2<0$ requires large and negative $H-\eta$ couplings which are in conflict with the stability conditions for the potential in this model.}.
We briefly summarize details on the different vacuum states, scalar masses, and stability conditions in App.~\ref{app:model}.
It is convenient to define the parameter: 
\begin{align}
R =\frac{\lambda_{345}}{\sqrt{\lambda_1\lambda_2}}~,~~	\lambda_{345}& = \lambda_3+\lambda_4 -|\lambda_5|~.
\label{rpara} 	
\end{align}
Finite-temperature corrections to the potential can be
parameterized by temperature-dependent masses to
leading order in the high-temperature limit
		\begin{align}
			\mu_H^2(T) = \mu_H^2 - c_1 T^2~,~~
			c_1&=\frac{3\lambda_1 +2\lambda_3 + \lambda_4}{12} + \frac{3 g^2 + g^{\prime2}}{32} + \frac{y_t^2}{8}\\ 
			\mu_\eta^2(T) = \mu_\eta^2 - c_2 T^2~,~~ 
			c_2&=\frac{3\lambda_2 +2\lambda_3 +\lambda_4}{12}+\frac{3g^2 + g^{\prime2}}{32}
	\end{align}
	where we neglected all other Yukawa couplings apart from the top quark Yukawa coupling $y_t$. From the stability conditions in eqs.~\eqref{eq:stability} we can infer that  $c_1 +c_2 >0$. We also restrict our discussion to the parameter space without charge-breaking vacuum states and thus impose $\lambda_4<|\lambda_5|$ and thus the VEVs are described by
	\begin{align}
	\braket{H} &=\frac{1}{\sqrt{2}} \begin{pmatrix} 0 \\ v_H \end{pmatrix}\;,\qquad
	\braket{\eta} = \frac{1}{\sqrt{2}}\begin{pmatrix} 0 \\ v_\eta  \end{pmatrix}\;.
\end{align}
Then there are four charge-conserving minima  
(where we indicate the VEVs in brackets): (i) the electroweak conserving minimum $EWc$ ($v_H=v_\eta=0$); (ii) the phenomenologically desired zero temperature minimum $I_1$ ($v_H\neq 0,~v_\eta=0 $), and two $Z_2$-breaking minima, (iii) $I_2$ ($v_\eta\neq 0,~v_H=0$), and (iv) $M$ ($v_\eta\neq0,v_H\neq0$).
	
\subsection{Thermal evolution}
\begin{figure}\centering
\includegraphics[width=1\linewidth]{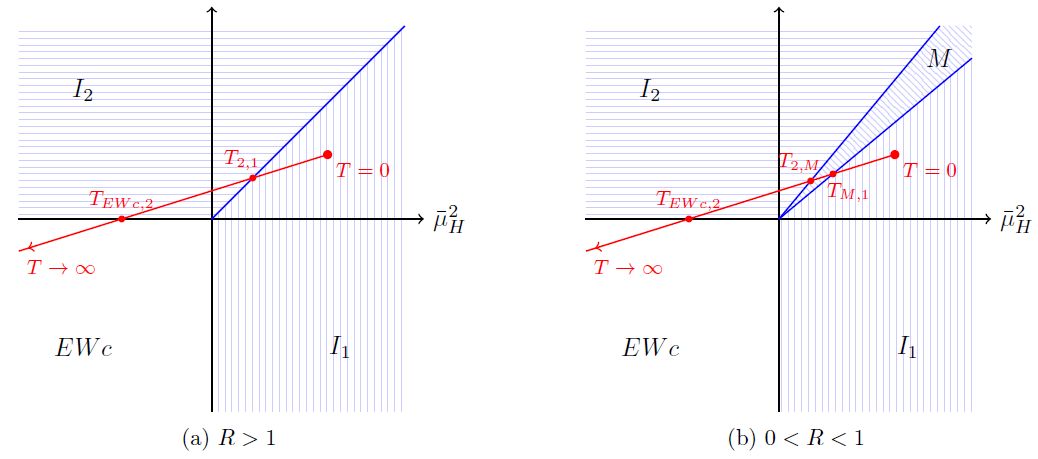}
\caption{Phase diagrams for different values of $R$. See Ref.~\cite{Ginzburg:2010wa} for a detailed discussion.}\label{fig:PhaseDiagram}
\end{figure}

The thermal evolution can be described by the trajectory in the plane $(\bar \mu_H^2(T),\bar \mu_\eta^2(T))$, where
\begin{align}
	\bar\mu_H^2(T)=\frac{\mu_H^2(T)}{\sqrt{\lambda_1}}~,~~
	\bar\mu_\eta^2(T)=\frac{\mu_\eta^2(T)}{\sqrt{\lambda_2}}~. 
\end{align}
We are interested in a thermal history which starts in the electroweak conserving phase $EWc$ with $\bar\mu_H^2(T)<0$ and $\bar\mu_\eta^2(T)<0$, undergoes a phase transition to one of the phases $I_2$/$M$ such that $\bar\mu_\eta^2(T)>0$ and ends in $I_1$ with $\bar\mu_H^2(T)>\bar\mu_\eta^2(T)$.
During the intermediate phase the discrete $Z_2$ symmetry is broken by $v_\eta\neq 0$ and finally restored after transitioning to $I_1$. Following Ref.~\cite{Ginzburg:2010wa} we identify two distinct thermal evolutions of our interest that take place for different values of parameter $R$ (\ref{rpara}) which are shown in Fig.~\ref{fig:PhaseDiagram}.

For $R>1$ there are two phase transitions with an
intermediate phase $I_2$ as shown in Fig.~\ref{fig:PhaseDiagram} (a). The first phase transition $EWc\to I_2$ occurs at $T_{EWc,2}$ while the second phase
transition $I_2\to I_1$ happens at $T_{2,1}$ and thus
the instability phase is given by the interval $[x_a, x_b]$ with: 
\begin{align}
	x_a  = \frac{M_N}{T_{EWc,2}} = \frac{\sqrt{c_2}M_N}{\mu_\eta}~,~~
	x_b  = \frac{M_N}{T_{2,1}} = \sqrt{\frac{\sqrt{\lambda_2}c_1 M_N^2- \sqrt{\lambda_1}c_2 M_N^2}{\sqrt{\lambda_2} \mu_H^2- \sqrt{\lambda_1}\mu_\eta^2}}~.
\end{align}
While the phase transition at $x_a$ is a second-order phase transition, the
second phase transition at $x_a$ is a first-order phase transition (at $x_b$
the two minima are degenerate and separated by a barrier). We assume the phase
transition to be instantaneous and estimate the entropy production $\Delta
s=Q_{I_2\to I_1}/T_b$ from the latent heat density $Q_{I_2\to I_1}$ released during
the phase transition~\cite{Ginzburg:2010wa}
\begin{equation}
	Q_{I_2\to I_1} = \left[T \frac{\partial V_{I_2}}{\partial T} - T \frac{\partial V_{I_1}}{\partial T}\right]_{T=T_b} 
	= \frac{c_1 \mu_\eta^2 - c_2 \mu_H^2}{4 \sqrt{\lambda_1\lambda_2}} T_b^2
\end{equation}
and thus $\Delta s= T_b (c_1 \mu_\eta^2 -c_2 \mu_H^2)/4\sqrt{\lambda_1\lambda_2}$.
The entropy dilution factor is then given by 
\begin{equation}
	\frac{S_f}{S_i} = 1 + \frac{\Delta s}{\frac{2\pi^2}{45} g_*^s(T_b) T_b^3} = 1+\frac{45}{8\pi^2\, \sqrt{\lambda_1\lambda_2}\,g_*^s(T_b)} \frac{c_1 \mu_\eta^2-c_2\mu_H^2}{T_b^2} 
	\simeq 1+0.006 \left(\frac{100}{g_*^s(T_b)}\right) \frac{c_1 \mu_\eta^2-c_2 \mu_H^2}{\sqrt{\lambda_1\lambda_2}T_b^2}\;,
\end{equation}
where $g_*^s(T_b)$ denotes the number of relativistic degrees of freedom (as they enter the entropy density) just before the phase transition at temperature $T_b$. 

For $0<R<1$ the phase diagram is more complicated as it contains three second-order phase transitions: $EWc\to I_2 \to M \to I_1$ for $0<R<1$, see Fig.~\ref{fig:PhaseDiagram} (b), with the corresponding critical temperatures $T_{EWc,2}$, $T_{2,M}$, and $T_{M,1}$, respectively.
The instability phase  is defined by: 
\begin{align}
	x_a & = \frac{M_N}{T_{EWc,2}} = \frac{\sqrt{c_2}M_N}{\mu_\eta}~,~~
	x_b  = \frac{M_N}{T_{M,1}} = \sqrt{\frac{ \lambda_{345} c_1 M_N^2 -  \lambda_1 c_2 M_N^2}{ \lambda_{345} \mu_H^2 - \lambda_1 \mu_\eta^2}}~.
\end{align}

\subsection{Evolution of DM abundance}
Having defined the above cosmological scenarios, we may proceed to calculate the DM abundance using the general formulae in sections \ref{sec:depop} and \ref{sec:regeneration}.  During the instability phase DM undergoes interactions through scattering and decay processes. The thermally averaged DM annihilation cross section is $p$-wave suppressed and can be approximated by
\begin{align}
	\braket{v\sigma (NN \to \nu\nu,\ell^+\ell^-)} \simeq \frac{T}{ 4\pi M_N^3} \left( \sum_{\alpha} |Y_{\alpha}|^2\right)^2~,
\end{align}
the scalar masses can be neglected compared to the DM mass \cite{Kubo:2006yx}. The DM decay ($N\to \eta L_a$) rate for heavy SM singlet fermions $M_N\gg m_{W,Z}$ is given by~\cite{Buchmuller:1991tu}
\begin{equation}
	\Gamma_{N}\simeq \frac{3 M_{N}}{32\sqrt{2}\pi}\frac{v_\eta^2}{v_\eta^2+v_H^2}\sum_\alpha |Y_{\alpha}|^2\;. 
\end{equation}
Note that there is an extra temperature dependence in the above equation which comes through thermal VEVs, $v_{\eta}$, $v_H$. This dependence can be neglected if $v_{\eta} \gg v_H$ and it drops out for the scenario with $R>1$ since $v_H=0$ during the instability phase.   

Next we turn to the regeneration phase, where $\eta$ particles produced during the instability phase decay back to DM particles, $\eta\to NL_a$, with width
\begin{align}
	\Gamma_{\eta}\simeq \frac{M_{\eta}}{16\pi}\sum_\alpha |Y_{\alpha}|^2 \left(1-\frac{M_N^2}{M_\eta^2}\right)^2\;. 
\end{align}
Large $M_{\eta}-M_N \gg M_{L_a}$ approximation was used in the above formula. The scalar mass $M_\eta$ weakly depends on the temperature $T$, but is well approximated by its zero temperature value and thus dominated by $\mu_\eta$. In addition, $\eta$ annihilates into electroweak gauge bosons. In the limit of $M_\eta\gg m_{W,Z}$ and neglecting the mass splitting induced by the VEVs we obtain for $s$-wave annihilation cross section of the electroweak doublet scalar $\eta$ the following decay width
\begin{align}
	\braket{v\sigma(\eta\eta^* \to BB,WW,B W)} & = \frac{|g^\prime|^4 + 3 |g|^4 + |g|^2|g^\prime|^2}{128\pi\, M_\eta^2}\;.
\end{align}

\subsection{Numerical Example}
\begin{figure}\centering
	\includegraphics[width=0.6\linewidth]{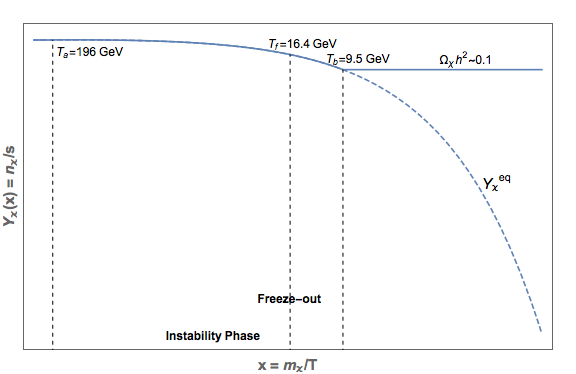}
	\caption{Plot of the DM abundance for the explicit example described in section \ref{subsec:sizeq} with the model parameters given above. After the phase transition ends, we approximately recover the observed relic abundance $(\Omega_{\rm DM} h^2) \sim 0.1$.}
	\label{fig2}
\end{figure}

As a proof of principle we give one numerical example which leads to the
correct DM relic density $(\Omega_{\rm DM} h^2)_{\text{obs.}} = 0.112 \pm
0.006$ \cite{refId0}. In particular, we examine the case explained in section \ref{subsec:sizeq} where the freeze out occurs during the phase transition such that the decays cause the dark matter to track the equilibrium yield until the phase transition ends. The quartic coupling $\lambda_1$ is fixed by the SM Higgs mass. In addition we choose the fermion mass $M_N$, the charged scalar mass
$M_{\eta^\pm}$, Yukawa couplings and the quartic couplings $\lambda_{2,3,4,5}$ as follows
\begin{align}
	M_{\eta^\pm} & = 264.6\,\mathrm{GeV} & 
	M_N & = 237.7\,\mathrm{GeV} &
	\sum_\alpha |Y_\alpha|^2 & = 1.7	\times 10^{-7}\\
	\lambda_2 & = 0.8294 &
	\lambda_3 & = 2.7709 &
	\lambda_4 & = 0.2153 &
	\lambda_5 & = -0.0303\;.
\end{align}
All other parameters are chosen at their respective best-fit values~\cite{Patrignani:2016xqp}. Then the phase transitions occur at $x_a=1.21$ ($T_a=196 \,\mathrm{GeV}$) and $x_b=25.1$ ($T_b=9.5\,\mathrm{GeV}$). The DM candidate $N$ is in kinetic and chemical equilibrium with the SM thermal plasma in the early universe and freezes out during the instability phase at $x_f=14.5$ ($T_f=16.4\,\mathrm{GeV}$), i.e. it is already non-relativistic at freeze-out. The entropy dilution factor from the second phase transition at $x_f$ is $S_f/S_i = 2.17$. The $W$, $Z$ and neutral scalar $\eta$ are non-relativistic at the end of the instability phase and thus their contributions to the DM abundance are exponentially suppressed. We show a plot of the evolution of the abundance in figure \ref{fig2}.

\newpage
\section{Discussion and conclusions}\label{sec:UVmodels}

In this paper we have proposed a mechanism of DM depopulation, which occurs due to the temporary breaking of a symmetry that ensures DM stability. During the instability phase the DM particles decay and their relative abundance drops down. Through these decays the primordial plasma gets also populated by additional species, which subsequently (after the instability phase) decay back to DM particles. The final DM abundance is established after this incomplete regeneration. 

We have demonstrated the DM depopulation mechanism using a simple extension of the standard model which contains a DM fermion $N$ and an additional electroweak doublet scalar field $\eta$, both of which are $Z_2$-odd. The latter field is assumed to be an inert field, i.e., without VEV at zero temperature. However, it plays a crucial role cosmologically in setting up the instability phase with broken $Z_2$ symmetry. With a suitable choice of parameters one can reduce an initially-large DM abundance down to the observed value. 

The fermionic DM model we have explicitly discussed has no deep theoretical motivation, but rather serves an illustrative purpose. The mechanism of depopulation can be applied to various theoretically better motivated models. In fact, our model is the closest analogue to the scotogenic model \cite{Ma:2006km} which incorporates neutrino masses together with fermionic DM within the SM  extension with two electroweak doublet scalars and three SM singlet fermions. Fermionic DM which is thermally produced via annihilations is in conflict with search for lepton-flavor-violating processes \cite{Vicente:2014wga}. Conversely, satisfying those constraints leads to low annihilation rates and thus overproduction of DM particles. The mechanism of depopulation can be straightforwardly applied to this class of models, resolving the tension with observations. 

Another class of well-motivated models with multiple scalar states are supersymmetric theories. The neutral lightest supersymmetric particle (LSP) (typically the neutralino) may serve as a natural candidate for cold DM in models with R-parity conservation. However, for significantly large space of parameters in supersymmetric extensions of the SM (especially for bino-like neutralinos) theory predicts DM to be over-abundant. The mechanism of depopulation can be applied to such models as well. Let us outline here one of the possible implementations of our mechanism within the minimal supersymmetric SM, where R-parity is broken temporarily during the instability phase of the sneutrino condensate. Imagine the relevant light states at low energies are just the SM Higgs boson and one sneutrino (the rest of sparticles are assumed to be heavy). The zero temperature scalar potential in this limit reads:  
\begin{equation}
V_0=-\frac{m_h^2}{4}h^2+m^2_{\tilde \nu}|\tilde \nu|^2+\frac{m_Z^2}{8v^2}\left(\frac{m_h}{m_Z}h^2+2|\tilde \nu|^2\right)^2~, \label{1}
\end{equation}
where $m_{\tilde \nu}$ is the sneutrino soft supersymmetry breaking mass parameter. Since we are interested in a phase where the sneutrino develops a non-zero vacuum expectation value, we allow it to be tachyonic\footnote{This assumption is atypical, yet phenomenologically viable. The tachyonic soft masses may emerge in some specific supersymmetry breaking scenarios at high energies.}. The physical sneutrino mass, however, receives contribution from the electroweak symmetry breaking and radiative corrections. The mass square, $M_{\tilde \nu}^2=m_{\tilde \nu}^2+\left(\frac{m_Zm_h}{2} + \text{rad. corr.}\right)$, must obviously be positive, hence the following constraint must be met: 
\begin{equation}
-\left(\frac{m_Zm_h}{2} + \text{rad. corr.}\right)< m_{\tilde \nu}^2< 0
\label{2}
\end{equation}

The dominant high temperature corrections to the potential (\ref{1}) reads (we ignore linear and logarithmic in $T$ terms):
\begin{eqnarray}
\label{3}
V_T&=& \frac{\alpha_hT^2}{2}h^2+\alpha_{\tilde \nu}T^2|\tilde \nu|^2~, \\
\label{4}
\alpha_h &=&\frac{1}{8v^2}\left(4m_W^2+2m_Z^2+4m_t^2+m_h^2+\frac{2}{3}m_Zm_h\right)\approx 0.383 \\
\label{5}
\alpha_{\tilde \nu} &=&\frac{1}{8v^2}\left(4m_W^2+4m_Z^2+\frac{1}{3}m_Zm_h\right)\approx 0.129~. 
\end{eqnarray}
The analysis of the full potential $V_0+V_T$ reveals that at sufficiently large $T$ the Higgs and the sneutrino fields reside in the origin, $\langle h \rangle_T=\langle \tilde \nu\rangle_T=0$, and hence electroweak symmetry and $R$-parity are exact.  Once the universe cools down to $T_c^{\tilde \nu}\approx 2.78|m_{\tilde \nu}|$, an instability develops in $\tilde \nu$-direction and the sneutrino field develops a non-zero vacuum expectation value, $\langle \tilde \nu\rangle_T=\frac{v}{m_Z}\left(-m_{\tilde \nu}^2-\alpha_{\tilde \nu}T^2\right)^{1/2}$, while the Higgs field remains at the origin. Further cooling down to $T_c^{h}\approx 143$ GeV results in the electroweak phase transition due to the Higgs field condensate $\langle h \rangle_T=\frac{v}{m_h}\left(m_h^2-2\alpha_h T\right)^{1/2}$. Positive contribution to the sneutrino mass parameter from the Higgs condensate starts to dominate and brings the sneutrino field back to the origin, restoring $R$-parity. Hence, for suitable $m_{\tilde \nu}^2$ we indeed account for a phase in the early universe,
\begin{equation}
T_c^h \approx 143~\text{GeV}~ < ~ T  <~ T_{c}^{\tilde \nu}\approx 2.78 |m_{\tilde\nu}|~,
\label{6}
\end{equation} 
where $R$-parity is broken spontaneously. During this phase the neutralino LSP ceases to be a stable particle. More specifically, condensation of the sneutrino field leads to a spontaneous breaking of $R$-parity and to a mixing of neutralinos and neutrinos. Through this mixing the neutralino LSP decays into standard model particles, the dominant decay channel being 2-body $\chi\to Z\nu'$ for $m_{\chi}>m_Z+m_{\nu}$. The longitudinal degrees of freedom of $Z$-boson during the instability state become massive sneutrino states and decay back to neutralino DM during the regeneration phase. Although the full phenomenological validity of this particular supersymmetric model requires further study, we are confident that the mechanism of depopulation can be implemented within supersymmetric models along the lines outlined here. 

In conclusion, we find that the proposed depopulation mechanism can be implemented within  various DM models to bring the DM abundance to observed valued, which otherwise would be considered empirically invalid.      
 
\paragraph{Acknowledgements} We would like to thank Lei Wu and Joachim Kopp for discussions. The work was partly supported by the Australian Research Council.

\appendix
\section{Details of the fermionic model}\label{app:model}
We briefly summarize the relevant details of the fermionic DM model.

\subsection{Scalar potential and symmetry breaking}
We restrict ourselves to the region of parameter space without a charge breaking vacuum, $\lambda_4<|\lambda_5|$. There is no vacuum state which spontaneously violates CP. Thus there are four relevant charge-conserving minima: The electroweak symmetry conserving minimum $EWc$, the desired zero temperature minimum $I_1$ where only the SM Higgs $H$ acquires a VEV, and the $Z_2$-breaking minima $I_2$ and $M$ which differ by the VEV of the SM Higgs $v_H$~\cite{Ginzburg:2007jn,Ginzburg:2010wa}:
\begin{align}
	EWc: & & v_H &= v_\eta =  0 & 
	V_{EWc} & =0\\
	I_1: && v_H^2 & = \frac{\mu_H(T)^2}{\lambda_1}, \quad v_\eta=0 &
	V_{I_1} & = - \frac{\mu_H(T)^4}{8\lambda_1}\\
	I_2: && v_\eta^2 &= \frac{\mu_\eta(T)^2}{\lambda_2}, \quad v_H=0 &
	V_{I_2} &  = -\frac{\mu_\eta(T)^4}{8\lambda_2}
\end{align}
\begin{align}
	M : & &
	v_H^2 &= \frac{\mu_H(T)^2\lambda_2 -\lambda_{345} \mu_\eta(T)^2}{\lambda_1\lambda_2 - \lambda_{345}^2}, \quad
	v_\eta^2 = \frac{\mu_\eta(T)^2\lambda_1 -\lambda_{345} \mu_H(T)^2}{\lambda_1\lambda_2 - \lambda_{345}^2} \\\nonumber &&
	V_M & = -\frac{\mu_H(T)^4\lambda_2 +\mu_\eta(T)^4 \lambda_1 -2\lambda_{345} \mu_H(T)^2\mu_\eta^2}{8(\lambda_1\lambda_2 -\lambda_{345}^2)}
\end{align}
We decompose the doublet fields in its components as follows
\begin{align}
	H&= \begin{pmatrix}
		h^+\\
		\frac{v_H + h + a}{\sqrt{2}}
	\end{pmatrix}&
	\eta&= \begin{pmatrix}
		\eta^+\\
		\frac{v_\eta + \eta_R + \eta_I}{\sqrt{2}}
	\end{pmatrix}\;.
\end{align}
Stability of the tree-level potential is ensured for~\cite{Deshpande:1977rw,Klimenko:1984qx,Nie:1998yn,Kanemura:1999xf,Ginzburg:2004vp,Kannike:2016fmd}
\begin{align}\label{eq:stability}
	\lambda_1 &>0 &
	\lambda_2 &>0 &
	\lambda_3 + \sqrt{\lambda_1\lambda_2} & >0 &
	\lambda_{345} +\sqrt{\lambda_1\lambda_2} & >0
\end{align}
The fourth stability condition implies the third one for $\lambda_4<|\lambda_5|$ and thus there are only $3$ conditions. Note that the last condition translates to $R>-1$. At zero temperature the minimum $I_1$ has to be the global minimum and $v_H^2 = 1/(\sqrt{2} G_F) \simeq (246\, \mathrm{GeV})^2$.

In the electroweak symmetry conserving vacuum, the masses of the components of $H$ and $\eta$ are given by
\begin{align}
	M_H^2 & = -\frac{\mu_H^2}{2} &
	M_\eta^2 & = -\frac{\mu_\eta^2}{2}\;. 
\end{align}
In the desired zero temperature vacuum with an unbroken discrete symmetry, $I_1$, the particle masses for the Higgs $h$ and the neutral ($\eta_{R,I}$) and charged ($\eta^\pm$) components of $\eta$ are
\begin{align}
	M_h^2 & = \lambda_1 v_H^2  &
	M_{\eta_R}^2 & = M_{\eta^\pm}^2 + \frac{\lambda_4+\lambda_5}{2} v_H^2 \\
	M_{\eta^\pm}^2 & = \frac{\lambda_3 v_H^2-\mu_\eta^2}{2} &
	M_{\eta_I}^2 & = M_{\eta^\pm}^2 + \frac{\lambda_4-\lambda_5}{2} v_H^2\;.
\end{align}
Note that $M_{\eta_I}>M_{\eta_R}$ for $\lambda_5<0$ and the lightest component of the scalar doublet $\eta$ is neutral for $\lambda_4<|\lambda_5|$.
In the vacuum $I_2$ where the discrete symmetry is broken, the masses are given by
\begin{align}
	M_{\eta_R}^2 & = \lambda_2 v_\eta^2&
	M_{h}^2 & = M_{h^\pm}^2 + \frac{\lambda_4+\lambda_5}{2} v_\eta^2 \\
	M_{h^\pm}^2 & = \frac{\lambda_3 v_\eta^2-\mu_H^2}{2} &
	M_{a}^2 & = M_{h^\pm}^2 + \frac{\lambda_4-\lambda_5}{2} v_\eta^2\;.
\end{align}
We do not report the masses for the vacuum $M$ where both VEVs, $v_H$ and $v_\eta$, are non-zero, because the expressions for the masses in the vacuum $M$ are more complicated and we not refer to them in the main part of the manuscript.
After electroweak symmetry breaking the masses of the electroweak vector bosons are given by
\begin{align}
	m_W & = \frac{g}{2} v &
	m_Z & = \frac{\sqrt{g^2+g^{\prime2}}}{2} v
\end{align}
with $v=\sqrt{v_\eta^2+v_H^2}$.

\subsection{Relevant cross sections and decay widths}
The DM annihilation cross section of fermionic DM $N$ into leptons is $p$-wave suppressed. Thus to leading order the annihilation cross section is given by~\cite{Kubo:2006yx}
\begin{align}
	v\sigma (NN \to \nu\nu,\ell^+\ell^-) = v^2 \frac{r^2 (1-2 r +2 r^2)}{24\pi M_N^2} \sum_{\alpha,\beta} |Y_{\alpha} Y_{\beta}^*|^2  \qquad
\text{with}\quad r=\frac{2 M_N^2}{M_{\eta_R}^2+M_{\eta_I}^2+ 2 M_N^2} 
\end{align}
with the relative velocity $v$. The thermally-averaged DM annihilation cross section is 
\begin{align}
	\braket{v\sigma (NN \to \nu\nu,\ell^+\ell^-)} = \frac{6T}{M_N} \frac{r^2 (1-2 r +2 r^2)}{24\pi M_N^2} \sum_{\alpha\beta} |Y_{\alpha} Y_{\beta}^*|^2 \;. 
\end{align}
For heavy DM $M_N\gg M_{\eta_R}, M_{\eta_I}$ during the instability phase, the DM annihilation cross section can be approximated by
\begin{align}
	\braket{v\sigma (NN \to \nu\nu,\ell^+\ell^-)} = \frac{T}{ 4\pi M_N^3} \left( \sum_{\alpha} |Y_{\alpha}|^2\right)^2. 
\end{align}
The decay rate of DM to leptons is~\cite{Buchmuller:1991tu}
\begin{align}
\Gamma_{N} & = \Gamma_{W^+\ell^-} + \Gamma_{W^-\ell^+} + \Gamma_{Z\nu}\\
\Gamma_{W^\pm\ell^\mp} & = \frac{G_F}{8\sqrt{2}\pi} |(U^\dagger\xi)_{\ell N}|^2 M_{N}^3 \left(1+2\frac{m_W^2}{M_{N}^2}\right)\left(1-\frac{m_W^2}{M_{N}^2}\right)^2\\
\Gamma_{Z\nu} & = \frac{G_F}{8\sqrt{2}\pi} |\xi_{\ell N}|^2 M_{N}^3 \left(1+2\frac{m_Z^2}{M_{N}^2}\right)\left(1-\frac{m_Z^2}{M_{N}^2}\right)^2
\end{align}
with the leptonic mixing matrix $U$ and the active-sterile mixing $\xi_{\ell N}\simeq \frac{Y_{\ell}v_\eta}{\sqrt{2} M_{N}}$ for vanishing final state lepton masses. Note that the Fermi constant in the $Z_2$-breaking phase is determined by $G_F^{-1}=\sqrt{2}(v_\eta^2 +v_H^2)$.
Summing over all flavors in the final state we find for $M_{N}\gg m_{W,Z}$
\begin{equation}
	\Gamma_{N}\simeq \frac{3 M_{N}}{32\sqrt{2}\pi}\frac{v_\eta^2}{v_\eta^2+v_H^2}\sum_\alpha |Y_{\alpha}|^2 \;. 
\end{equation}


\bibliographystyle{h-physrev}
\bibliography{wimp}

\end{document}